\documentclass[iop]{emulateapj}
\usepackage{graphicx}
\usepackage{natbib}
\usepackage{ulem}
\usepackage[usenames]{color}
\usepackage{epstopdf}
\usepackage{enumitem}
\slugcomment{~~\today}

\newcommand       \clash        {CLASH~2882}
\newcommand       \msil        {$M_{sil}$}
\newcommand       \mcrb        {$M_{crb}$}
\newcommand       \tsil        {$T_{sil}$}
\newcommand       \tcrb        {$T_{crb}$}
\newcommand       \rsn         {$R_{\rm SN}$} 
\newcommand       \rccsn         {$R_{\rm CCSN}$}
\newcommand       \ragb         {$R_{\rm AGB}$}  
\newcommand       \rsne         {R_{\small \rm SN}} 
\newcommand       \rccsne         {R_{\small \rm CCSN}}
\newcommand       \ragbe         {R_{\rm AGB}} 
\newcommand       \mism         {$M_{\rm ISM}$} 
\newcommand       \misme         {M_{ \rm  ISM}}
\newcommand       \mul         {$\mu^{-1}$} 
\newcommand       \msun        	{$M_{\odot}$}
\newcommand       \lsun      	{$L_{\odot}$} 
\newcommand       \hub {km~s$^{-1}$~Mpc$^{-1}$}

\newcommand	      \sfr              {$M_{\odot}$~yr$^{-1}$}
\newcommand        \mic        	 {$\mu$m}

\newcommand   \chisq     {$\chi^2$}

\newcommand      \hst   {{\it HST}}
\begin{document}

\title{Submillimeter Observations of CLASH~2882 and the Evolution of Dust in this Galaxy}

\author{Eli Dwek\altaffilmark{1},Johannes Staguhn\altaffilmark{1,2}, Richard G. Arendt\altaffilmark{1,3}, Attila Kov\'{a}cs\altaffilmark{4}, Roberto Decarli\altaffilmark{5},  Eiichi Egami\altaffilmark{6}, , Micha{\l} J.~Micha{\l}owski\altaffilmark{7}, Timothy D. Rawle \altaffilmark{8,9}, Sune Toft\altaffilmark{10}, and Fabian Walter \altaffilmark{5} }
\affil{Observational Cosmology Lab., Code 665 \\ NASA Goddard Space Flight Center,
Greenbelt, MD 20771, \\ e-mail: eli.dwek@nasa.gov}
\altaffiltext{1}{Observational Cosmology Lab., Code 665, NASA at Goddard Space Flight Center, Greenbelt, MD 20771, USA}
\altaffiltext{2}{The Henry A. Rowland Department of Physics and Astronomy, Johns Hopkins University, 3400 N. Charles Street, Baltimore, MD 21218, USA}
\altaffiltext{3}{CRESST, University of Maryland Baltimore County, Baltimore, MD 21250, USA}
\altaffiltext{4}{Astronomy Department, CalTech, Pasadena, CA 90025 and Astronomy Department, University of Minnesota, MN 12345, USA}
\altaffiltext{5}{Max-Planck Institut fuer Astronomie Koenigstuhl 17, 69117 Heidelberg, Germany}
\altaffiltext{6}{Steward Observatory, University of Arizona
933 N. Cherry Ave., Tucson AZ 85721, USA}
\altaffiltext{7}{SUPA (Scottish Universities Physics Alliance), Institute for
Astronomy, University of Edinburgh, Royal Observatory, Edinburgh, EH9
3HJ, UK}
\altaffiltext{8}{European Space Astronomy Centre (ESA/ESAC), E-28691 Villanueva de la Ca\~{n}ada, Madrid, Spain}
\altaffiltext{9}{ESA/STScI, 3700 San Martin Drive, Baltimore, MD 21218, USA}
\altaffiltext{10}{Dark Cosmology Centre, Niels Bohr Institute, Copenhagen, Denmark}
\setcounter{footnote}{10}
\begin{abstract}
Two millimeter observations of the MACS~J1149.6+2223 cluster have detected a source that was consistent with the location of the lensed MACS1149-JD galaxy at $z=9.6$. A positive identification would have rendered this galaxy as the youngest dust forming galaxy in the universe. Follow up observation with the AzTEC 1.1~mm camera and the IRAM NOrthern Extended Millimeter Array (NOEMA) at 1.3~mm have not confirmed this association. In this paper we show that the NOEMA observations associate the 2~mm source with [PCB2012]~2882\footnote{[PCB2012] 2882 is the NED-searchable name for this source.}, source number 2882 in the {\it Hubble Space Telescope} ({\it HST}) Cluster Lensing and Supernova (CLASH)  catalog of MACS~J1149.6+2223. This source, hereafter referred to as \clash, is a gravitationally lensed spiral galaxy at $z=0.99$. We combine the GISMO 2~mm and NOEMA 1.3~mm fluxes with other (rest frame) UV to far-IR observations to construct the full spectral energy distribution (SED) of this galaxy, and derive its star formation history,  and stellar and interstellar dust content. The current star formation rate of the galaxy is 54$\mu^{-1}$~\sfr, and its  dust mass is about $5\times 10^7$\mul~\msun, where $\mu$ is the lensing magnification factor for this source, which has a mean value of 2.7. The inferred dust mass is higher than the maximum dust mass that can be produced by core collapse supernovae (CCSN) and evolved AGB stars.  As with many other star forming galaxies, most of the dust mass in \clash\ must have been accreted in the dense phases of the ISM.
\end{abstract}

\section{INTRODUCTION}
Deep 2~mm observations of the MACS~J1149.6+2223 cluster field with the Goddard IRAM 2 Millimeter Observer (GISMO) revealed a 2~mm source \citep{staguhn14} that was consistent with the position of the gravitationally lensed galaxy MACS1149-JD located at $z=9.6$ \citep{zheng12}.  
Assuming the validity of this association, \cite{dwek14} analyzed the dust formation and destruction rates that are unique to the very high redshift universe. An important general result of their study was that in high redshift galaxies with dust-to-gas mass ratios below a critical value of $\sim 10^{-4}$, and hence metallicities below $\sim 3\times10^{-4}$, supernova are net producers of interstellar dust, so that the net rate of dust formation in such galaxies exceeds that in older, more metal-rich, objects.

However, imaging from the Herschel Lensing Survey \citep{egami10} indicated that the FIR flux may instead originate from one of two galaxies at $z\sim 1$, also within the GISMO 17.5\arcsec\ beam [see cautionary note \cite{dwek14}].
Follow up observations  with the AzTEC 1.1~mm camera mounted on the Large Millimeter Telescope {\it Alfonso Serrano}, provided an image of the MACS1149-JD field with 8.5\arcsec\ resolution \cite{zavala15}. Their observations detected a 3.5$\sigma$ source consistent with the GISMO position, associated with a group of galaxies located 11\arcsec\ away from MACS1149-JD source. Five sources with redshifts between 0.7-1.6, that were detected in CLASH survey of this field \citep{postman12}, are within the AzTEC beam, preventing the definitive association of the GISMO source with an individual galaxy.

In this paper we present a 1.3\arcsec\ resolution image of the 33\arcsec\ $\times$ 33\arcsec\ field centered around MACS1149-JD  obtained at 1.3~mm with the NOEMA. The image shows a 4$\sigma$ source, located within the AzTEC beam, that is positively identified with \clash, a galaxy at a redshift of 0.99.      
We combine the GISMO 2~mm observations with the UV-optical (UVO) to far-infrared (IR) and submillimeter observations to construct the  spectral energy distribution (SED) of the stellar and dust emission components of the galaxy. 

This paper is organized as follows. The observations are presented in Section~2. The galaxy's SED is used to derive its dust mass and possible composition, and its current star formation rate (SFR) and possible star formation history (SFH). The results are presented in Section~3.  In Section~4 we discuss the maximum attainable dust mass from stellar sources alone, and compare it with the dust masses inferred from the observations. The origin of the dust and a brief summary of the paper are presented in Sections 5 and 6, respectively. 

 In all our calculations we adopt a Hubble constant of 70~\hub, and values of $\Omega_m = 0.27$, and $\Omega_{\Lambda}=0.73$ for the critical densities of dark matter and dark energy, respectively \citep{hinshaw09}. The age of \clash, taken to be at $z=1$, is  6~Gyr, its distance is 6,750~Mpc, and its angular diameter distance, defined as the ratio between the galaxy's transverse size to its angular size (in radians), is 1,700~Mpc.      
    
 \section{Observations of CLASH~2882}

Given the importance of identifying the counterpart to the 2~mm GISMO source, we targeted the field of MACS1149-JD using the NOEMA on December 30, 2014, and January 1, 2015 (program: W14FP, PI: Staguhn). The array operated in compact (7D) configuration. The tuning frequency (231.86 GHz) was chosen to encompass the [NII] 122\mic\ line in the redshift range $9.50<z<9.67$. The pointing center was set on the coordinates of the NIR emission of MACS1149-JD \citep{zheng12}. At the observing frequency, the primary beam of NOEMA is 21.7\arcsec. The data processing was performed with the latest version of the GILDAS suite, in particular with CLIC for the calibration and flagging, and with MAPPING for the imaging. The final data cube consists of 8784 visibilities (7.32 hr of integration, 6-antennas equivalent). We reached a sensitivity of 1.05 mJy/beam per 50 km/s wide channel, or 103 $\mu$Jy/beam in the collapsed continuum map (at 1-$\sigma$). The synthesized beam is 2.2\arcsec$\times$1.5\arcsec, PA=$11^\circ$.
 
\begin{figure}[tbp] 
   \centering
   \includegraphics[width=1.5in]{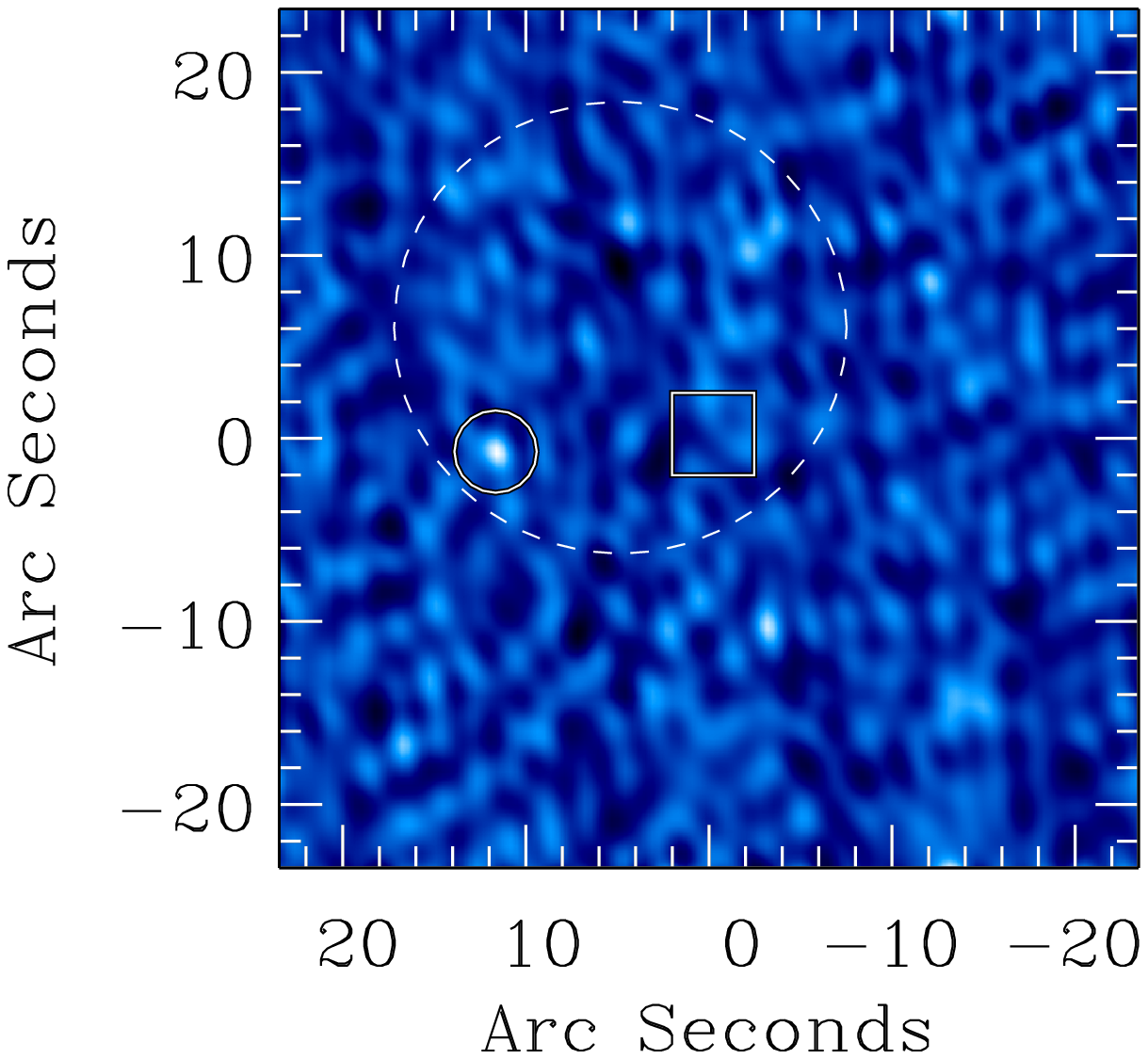} 
   \includegraphics[width=1.5in]{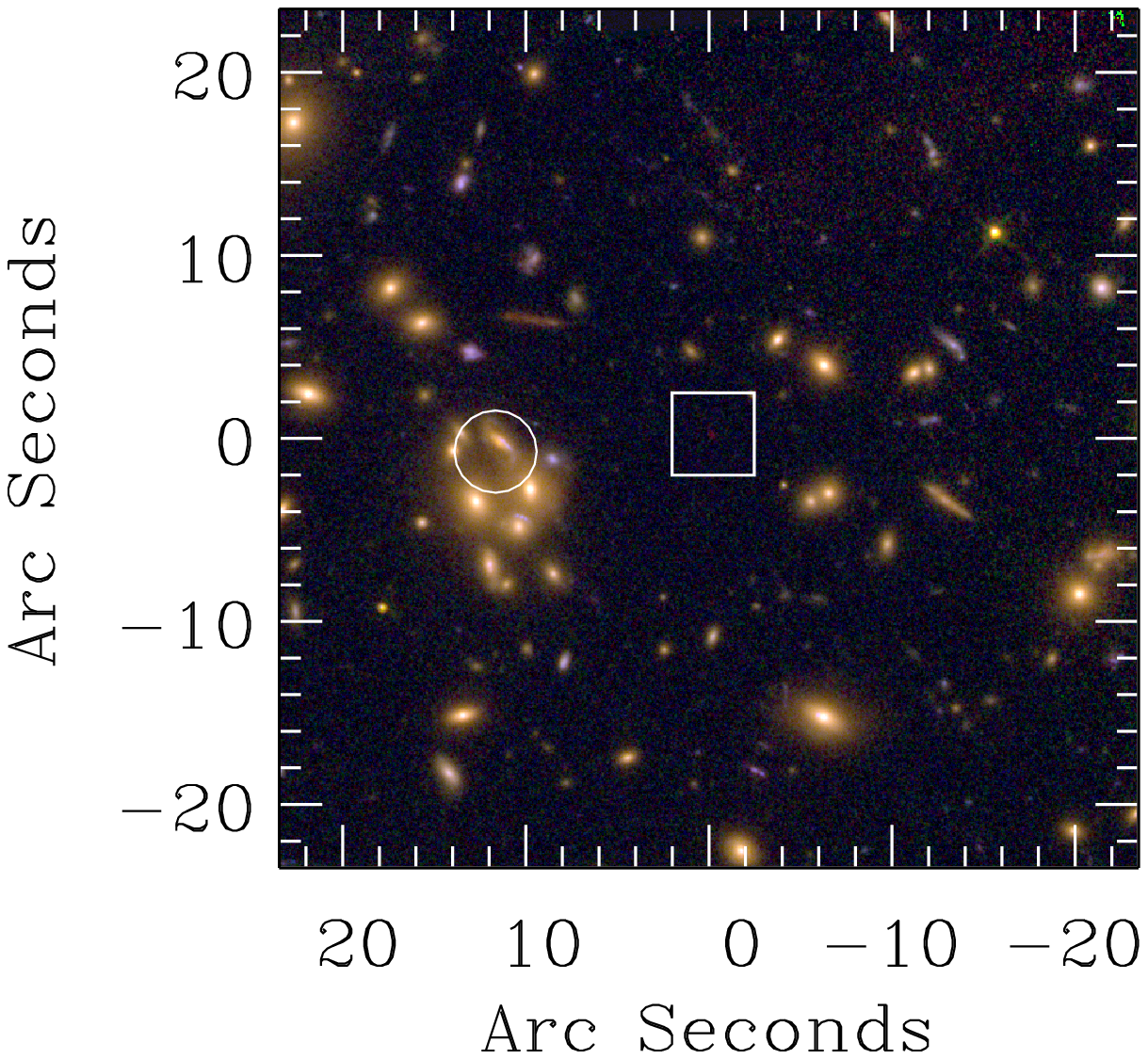}\\
   \includegraphics[width=1.5in]{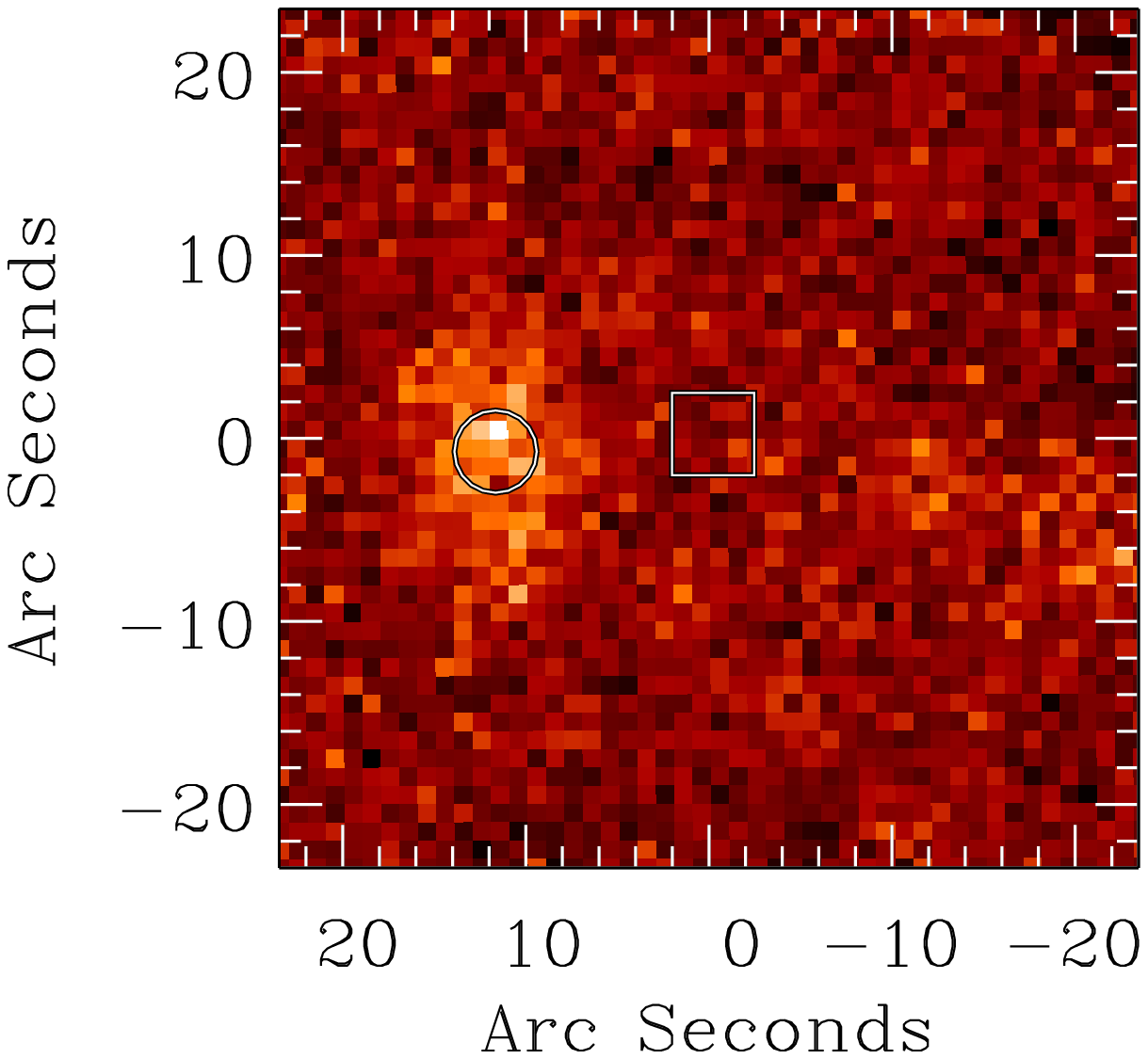} 
   \includegraphics[width=1.5in]{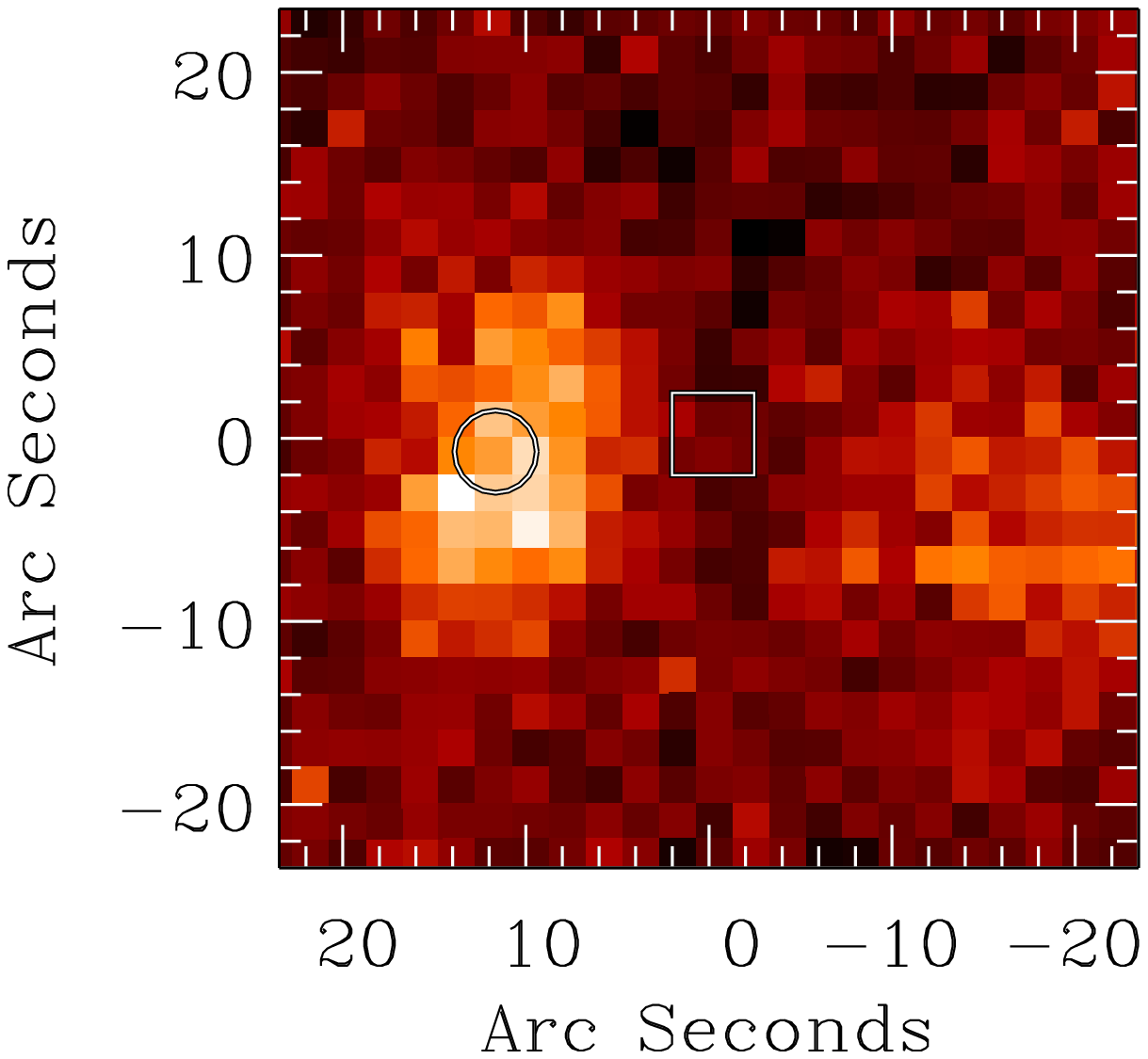}\\ 
   \includegraphics[width=1.5in]{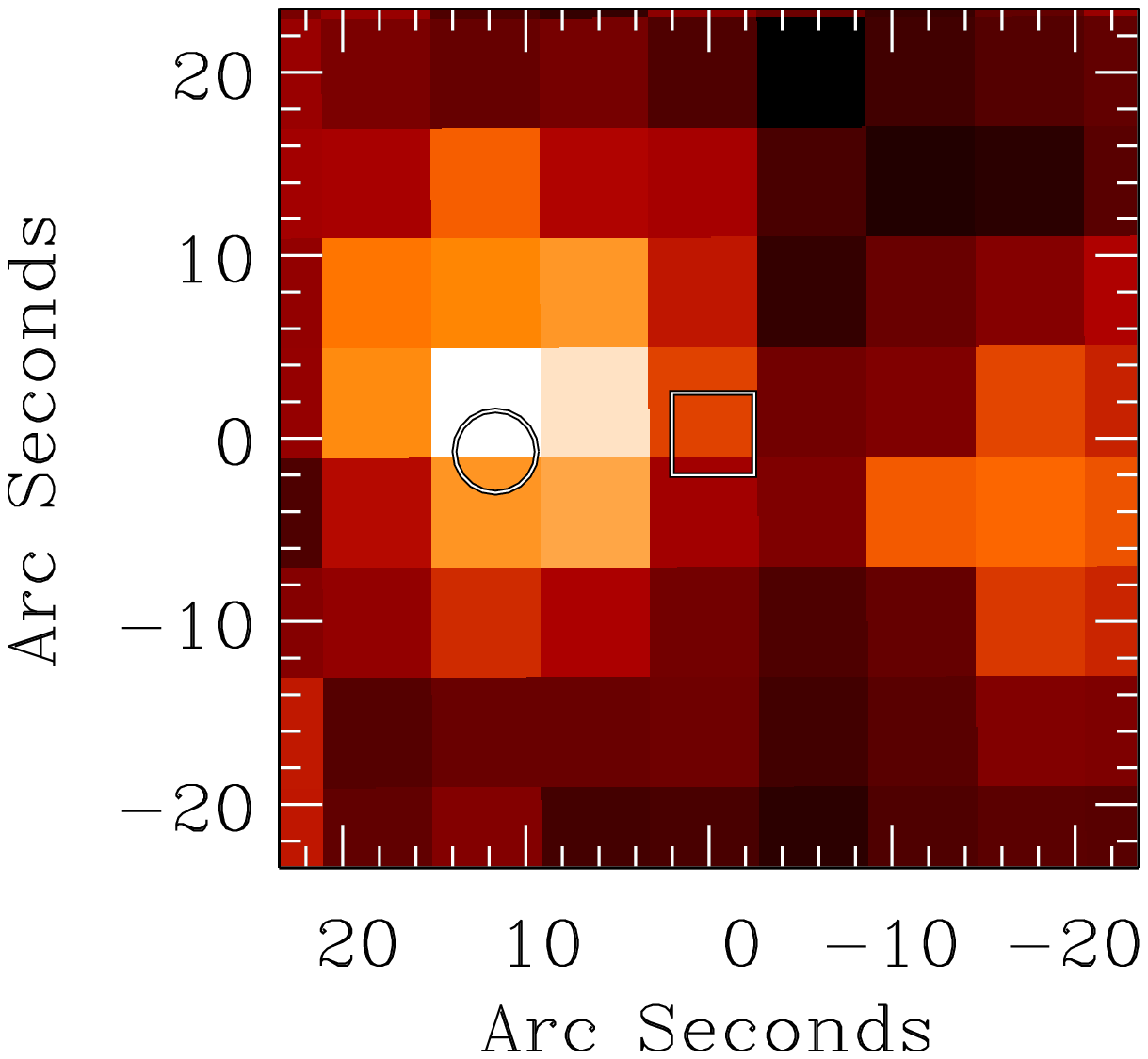} 
   \includegraphics[width=1.5in]{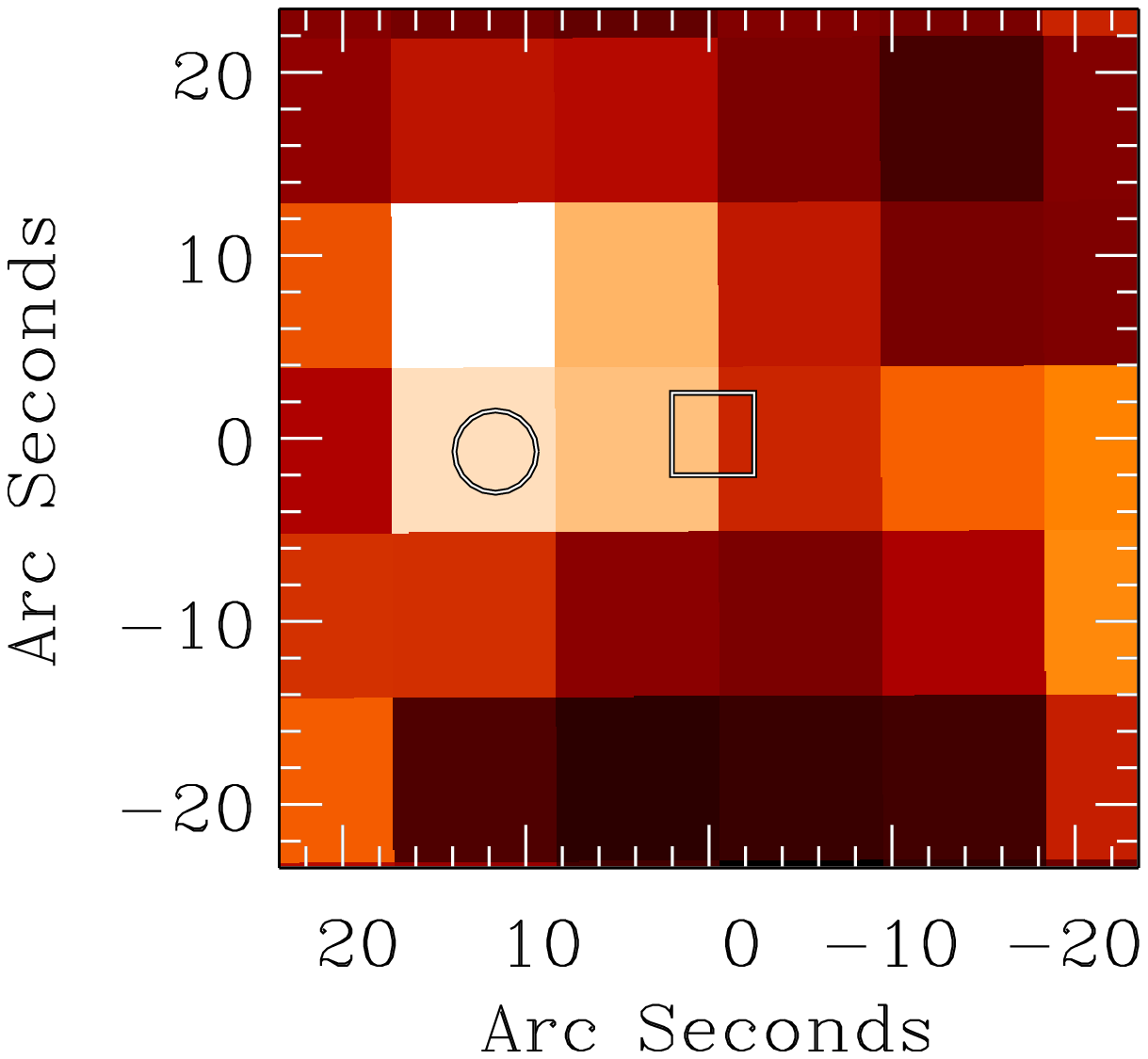}
   \caption{Images of CLASH 2882 and MACSJ1149-JD. From left to right and top to bottom the images represent: (a) the NOEMA 1.3~mm image.
   The location of the 4$\sigma$ source we associate with CLASH 2882 is marked
   with the small circle. The square marks MACSJ1149-JD. The large dashed 
   circle indicates the GISMO 2~mm effective FWHM, centered at the nominal 
   location of the GISMO source. (b) {\it HST} image at 1.60, 1.05, 0.555 \mic.
   (c-f) {\it Herschel} PACS and SPIRE images at 100, 160, 250, and 350 \mic.
   All images are centered at $(\alpha,\delta) = (11:49:33.60,+22:24:45.5)$.\label{clash2882}} 
 \end{figure}

The image of the field, centered on the MACSJ1149-JD source, is presented in Figure~\ref{clash2882}a. MACSJ1149-JD is marked with a square. The most prominent source in the field (marked by a circle) is associated with \clash. Figure~\ref{clash2882}b is the \hst\ image of the same field at 1.60, 1.05, and 0.555~\mic. Figures~\ref{clash2882}c-f, show the {\it Herschel} PACS and SPIRE images at 100, 160, 250, and 350~\mic, respectively, observed as part of the {\it Herschel} Lensing Survey \citep{egami10} (Rawle 2015, in preparation).

\begin{figure}[htbp] 
   \centering
   \includegraphics[width=3.5in]{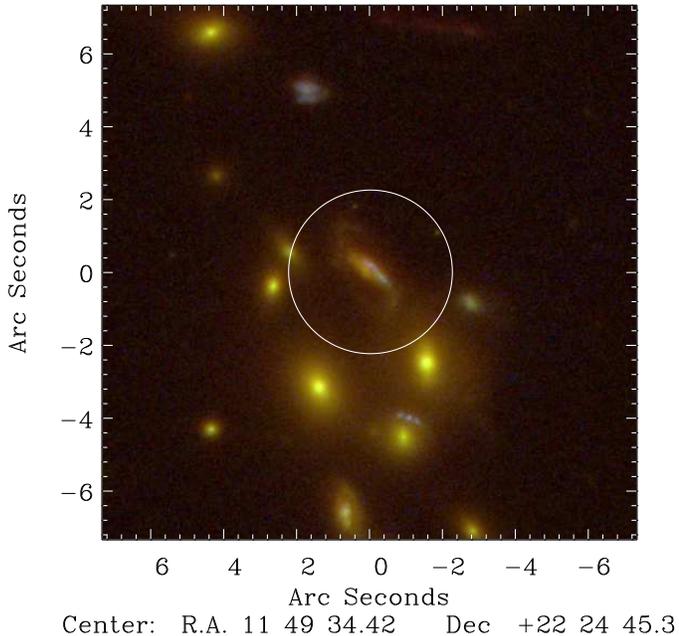}  
    \caption{Detailed {\it HST} image of [PDB2012] 2882 (circled) in the 
F110W, F814W, and F435W filters from the CLASH observations.\label{clashmap}} 
 \end{figure}

Figure~\ref{clashmap} is a more detailed view of the \hst\ image of \clash. The galaxy appears redder than the other sources in the group. The map also shows a population of blue stars at the south-western end of this galaxy, representing a region of unobscured star formation that may, or may not, be associated with \clash. 

\setcounter{footnote}{11}
\clash\ is lensed by the MACSJ1149.5+2223 cluster at $z=0.543$. Using the Frontier Fields Lens Models\footnote{https://archive.stsci.edu/prepds/frontier/lensmodels/}
we find that the median lensing amplification factor ranges from 1.61 to 4.21, with a average value of $\mu =2.70$.

\begin{figure}[htbp] 
   \centering
   \includegraphics[width=3.0in]{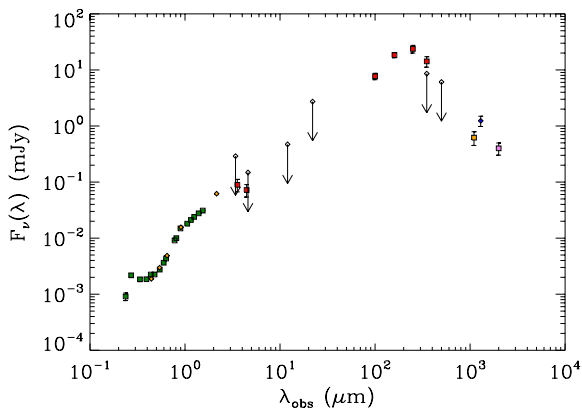} 
    \caption{The observed flux densities from \clash. The SED comprises of contributions from an unobscured star forming regions ($\sim 0.2-0.4$~\mic); an obscured stellar population ($\sim 0.4-2.0$~\mic); and thermal emission from dust ($\gtrsim 50$~\mic). These components are not evident here, but are shown in Fig. \ref{specfits}. \label{clashsed}} 
 \end{figure}

\begin{figure*}[htbp] 
   \centering
 \includegraphics[width=2.5in]{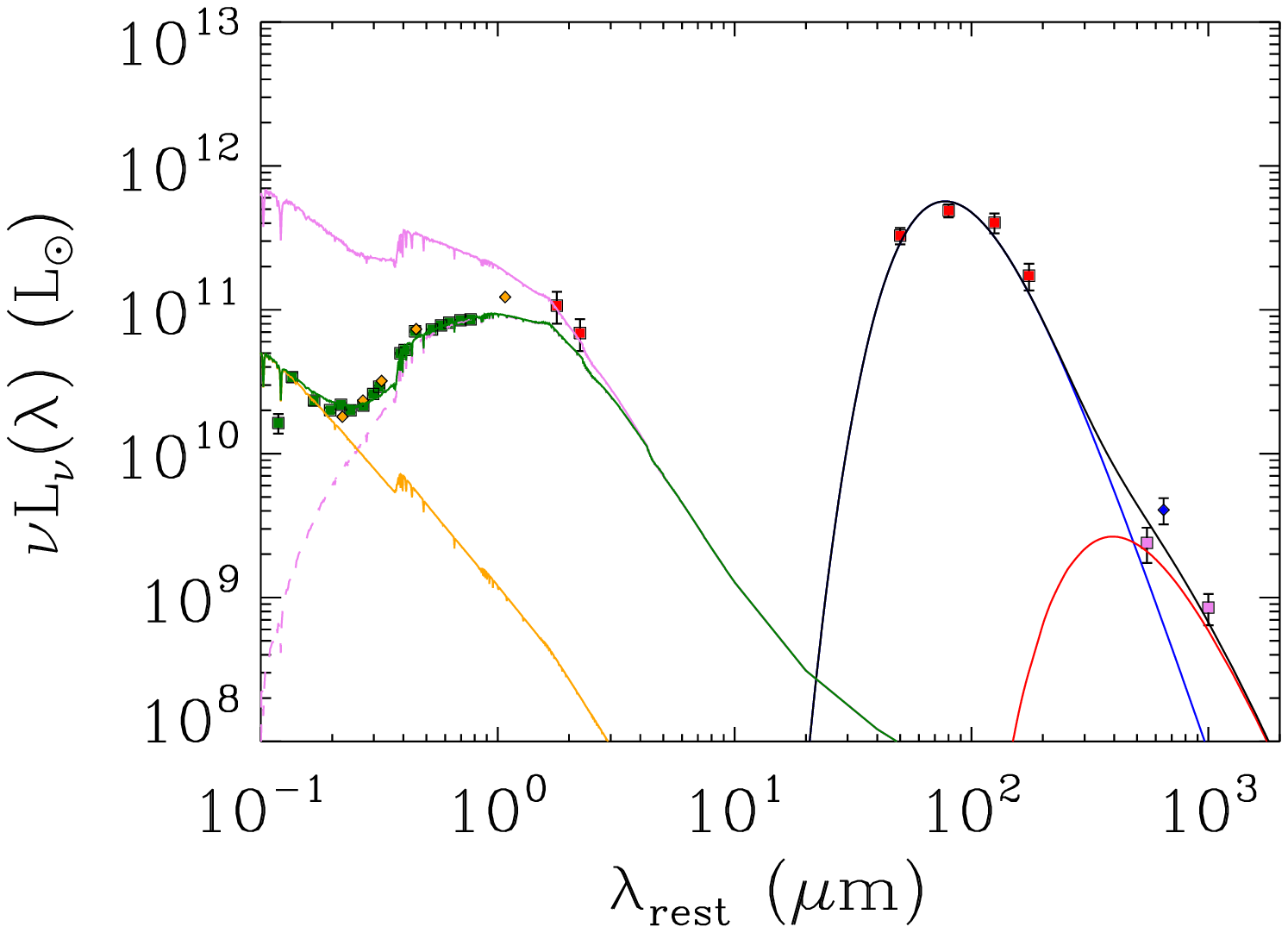} ~~~~~
   \includegraphics[width=2.5in]{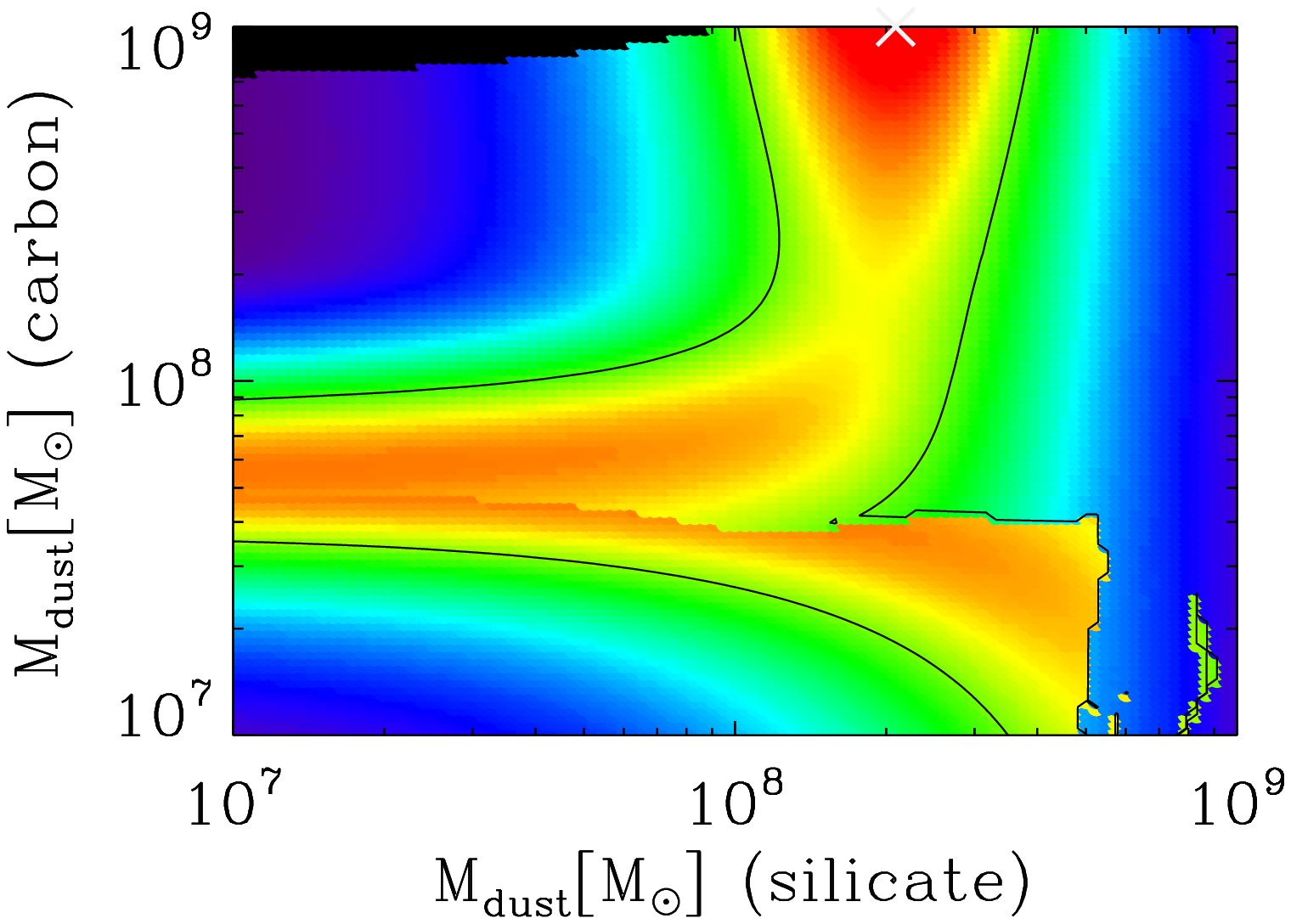} \\
      \includegraphics[width=2.5in]{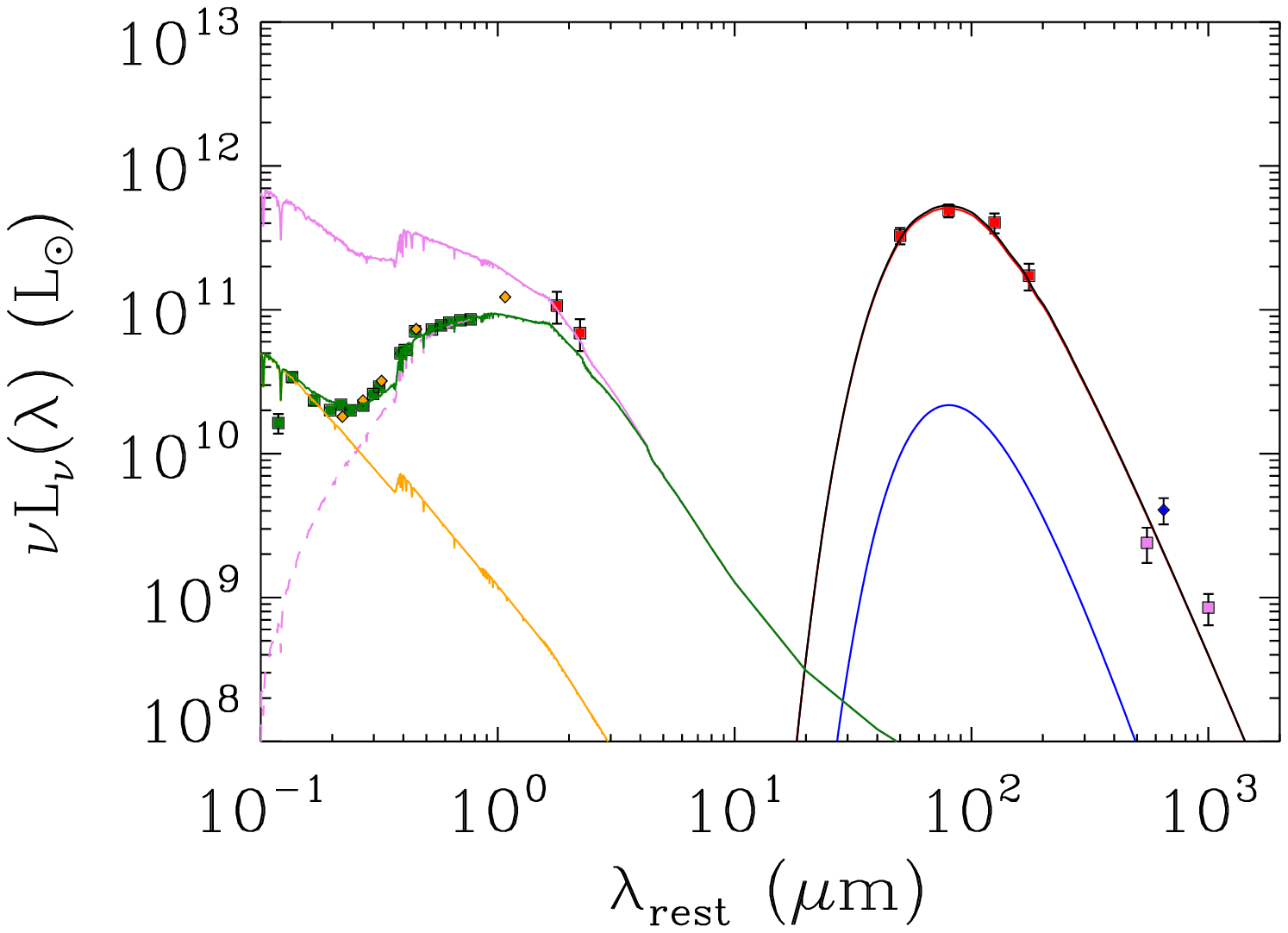} ~~~~~ 
      \includegraphics[width=2.5in]{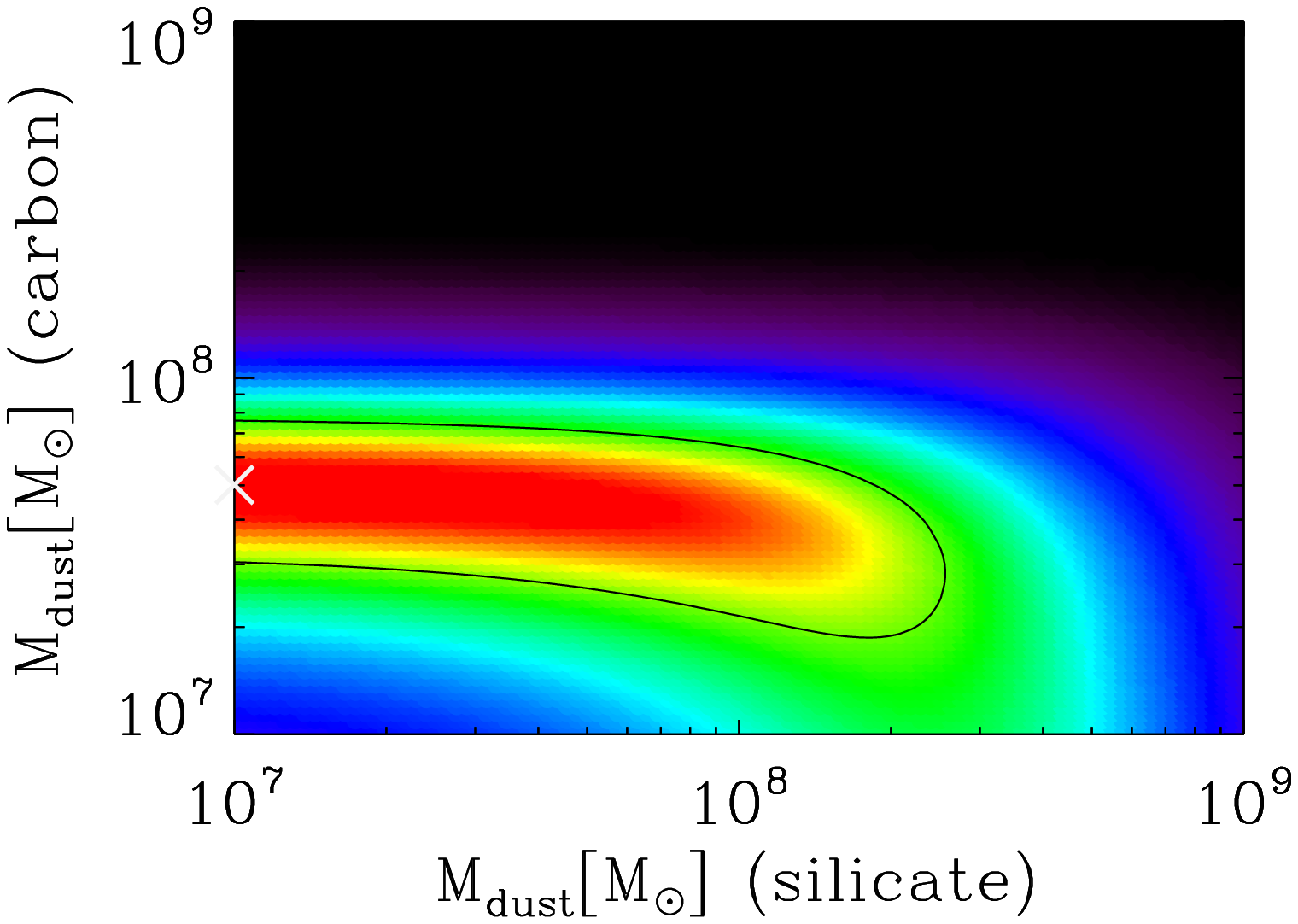} \\
         \includegraphics[width=2.5in]{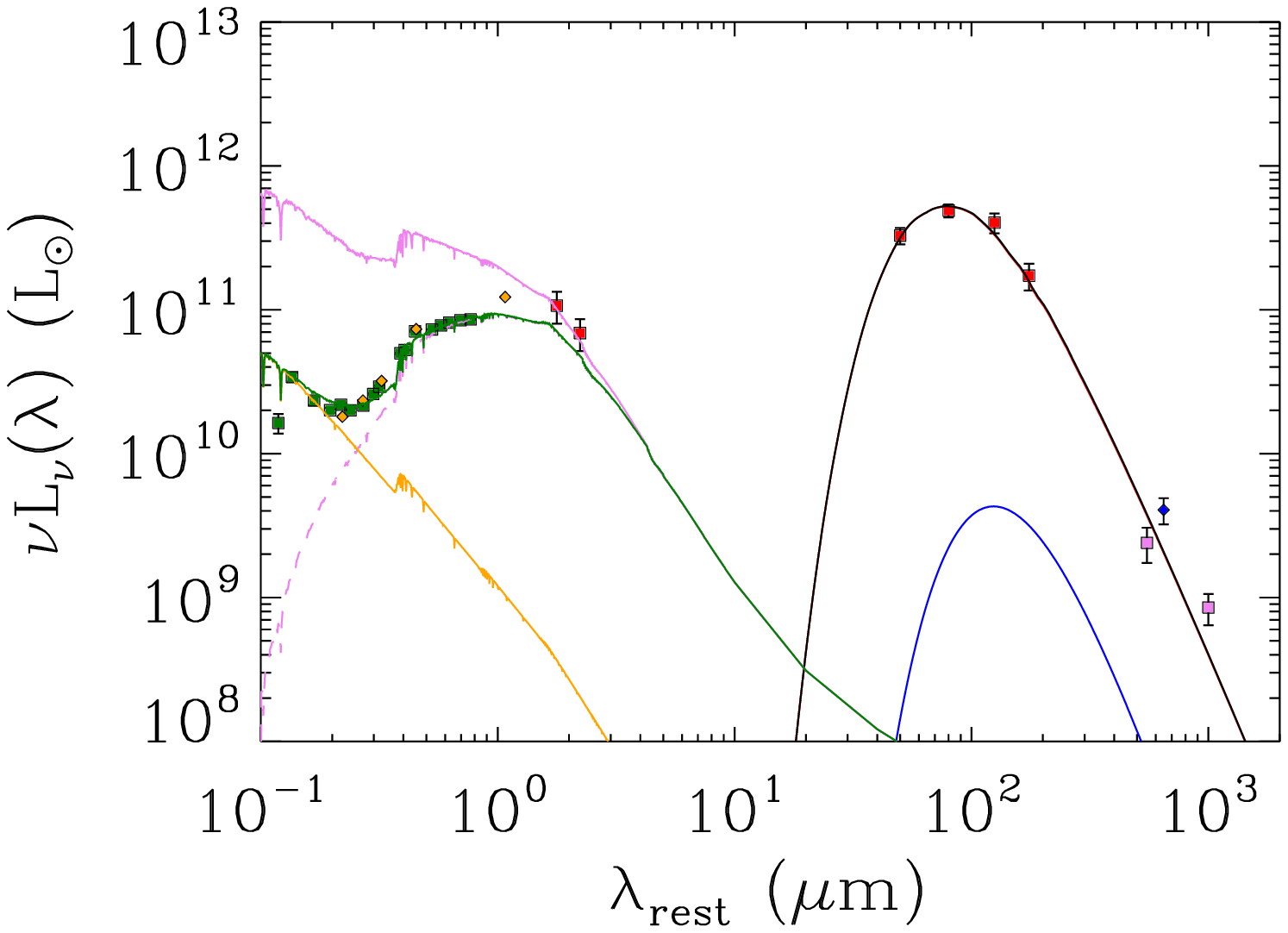}  ~~~~~
         \includegraphics[width=2.5in]{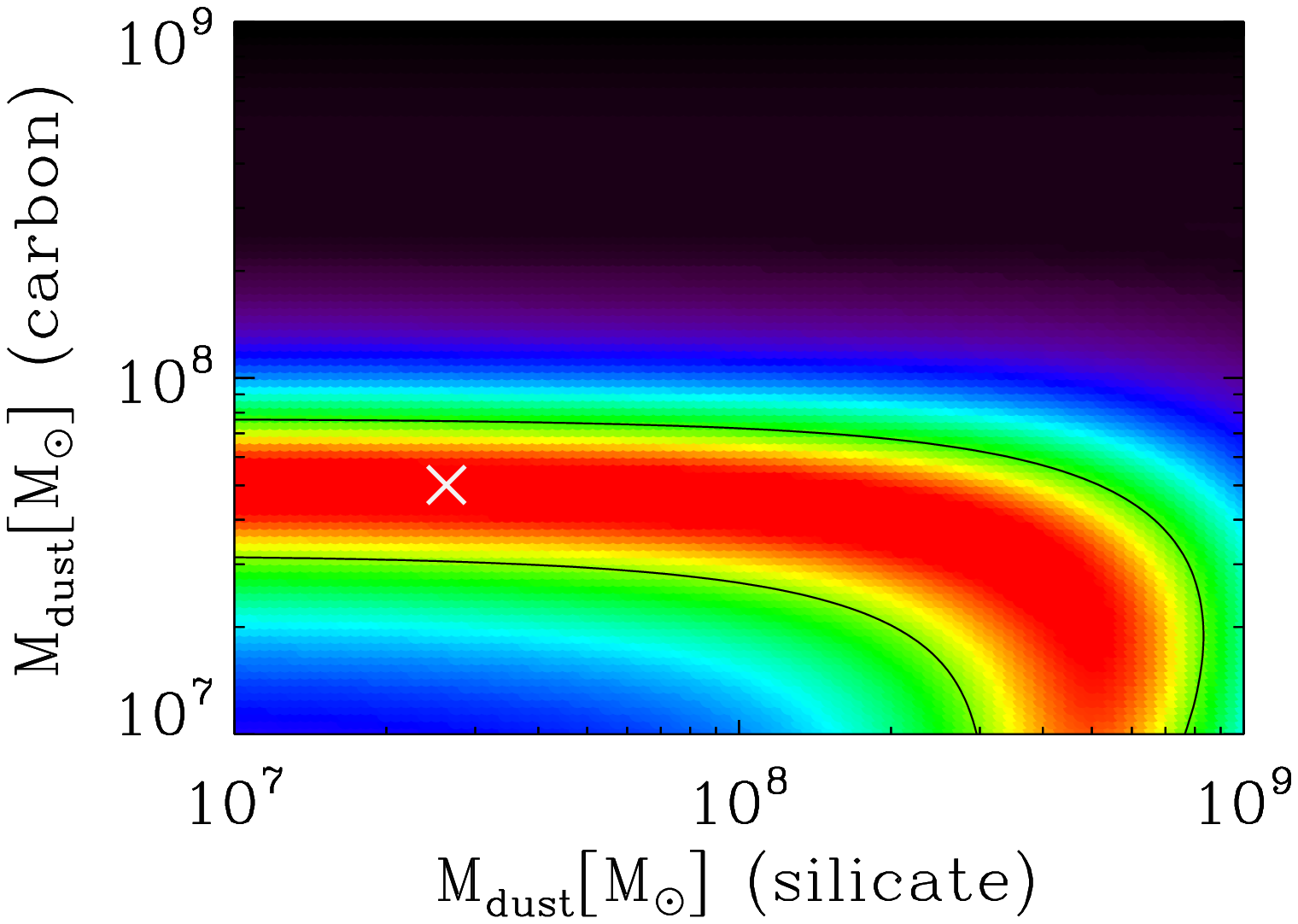} \\
            \includegraphics[width=2.5in]{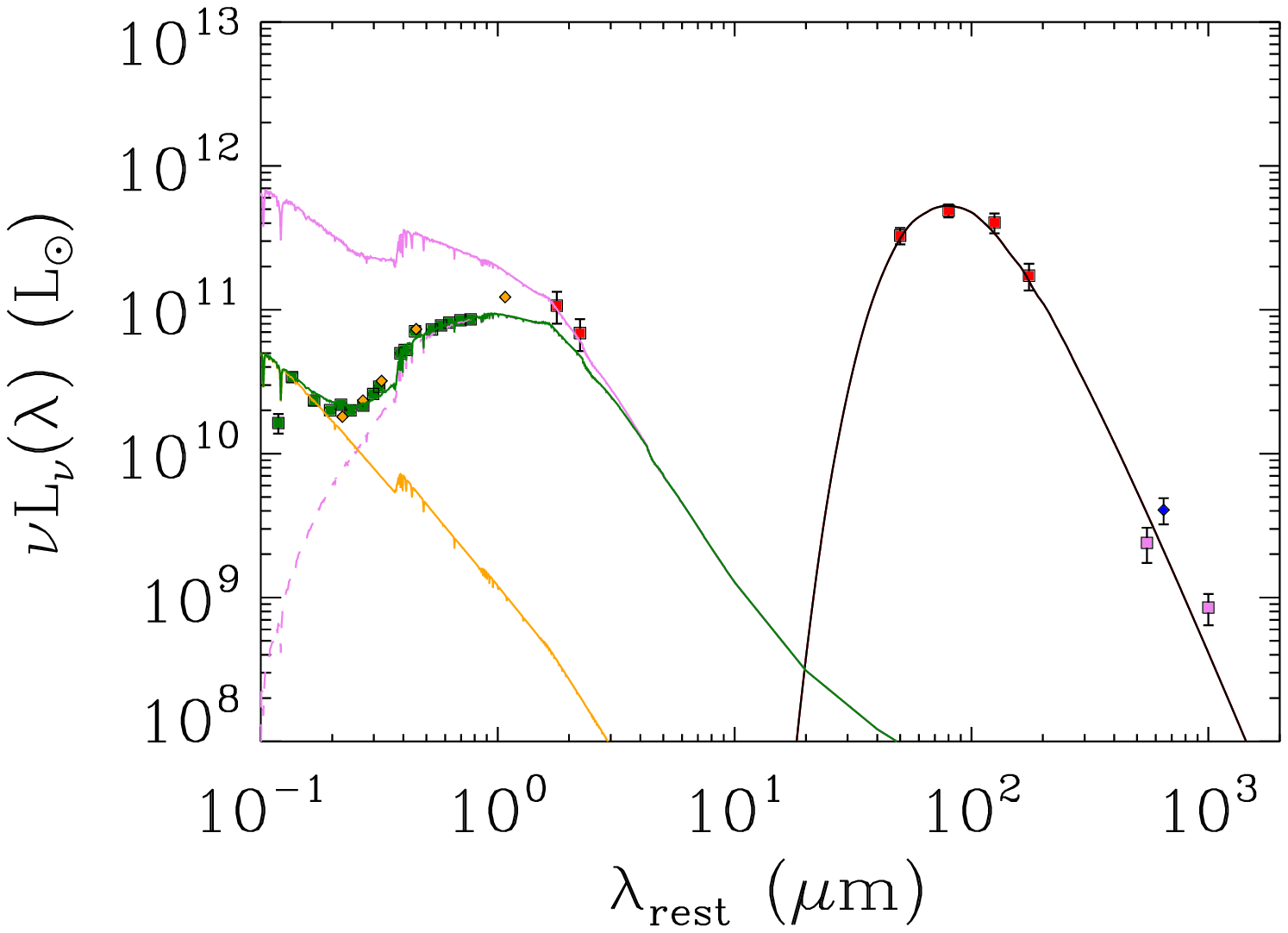}  ~~~~~
            \includegraphics[width=2.5in]{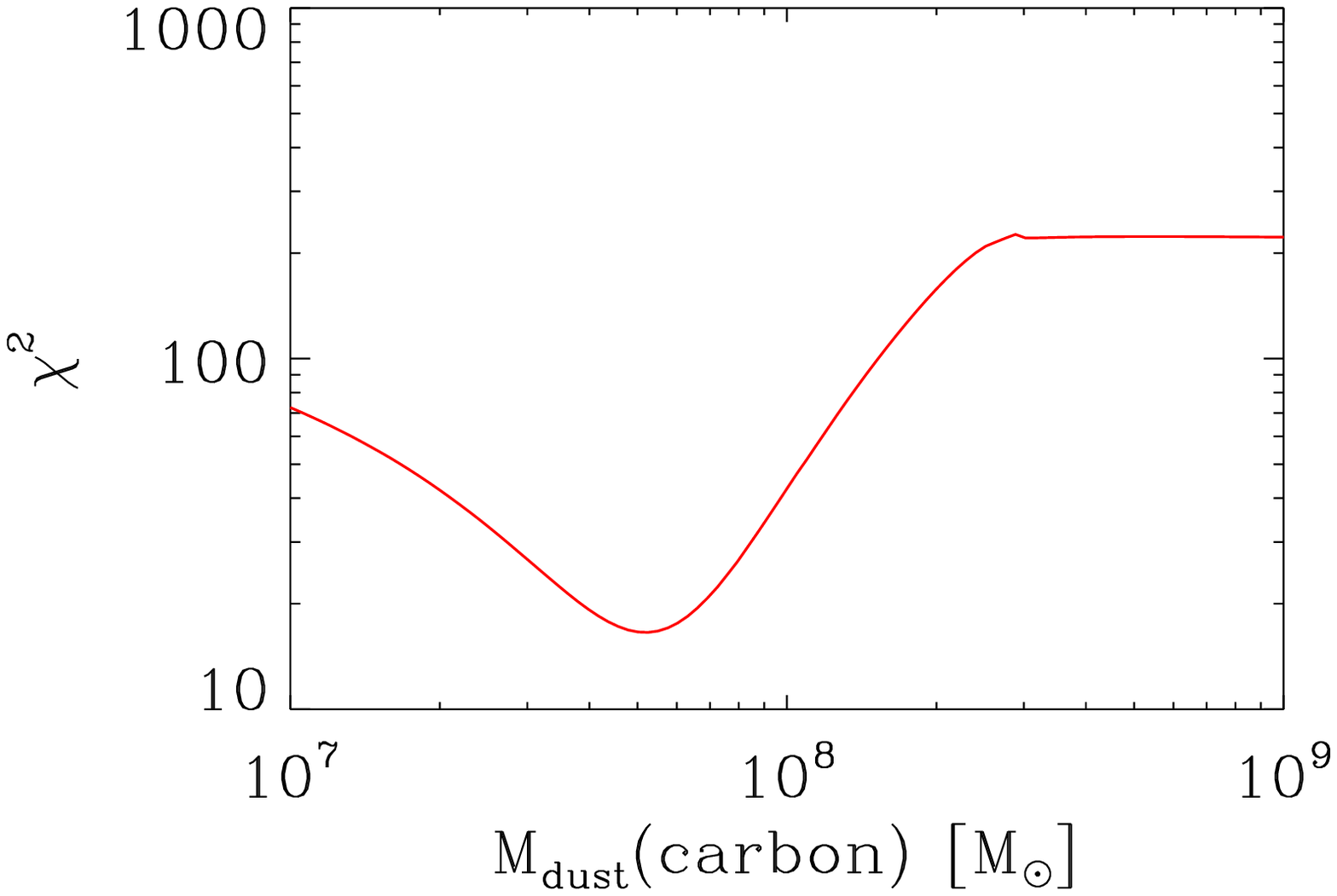} \\ 
   \caption{ {\bf Left column}: The best fits to the stellar and dust emission spectra. The intrinsic and escaping stellar spectra are shown in solid and dashed violet, respectively. The orange curve represents the spectrum from the starburst population, and the green curves represent the sum of the escaping stellar spectra. The blue and red curves represent the silicate and AC dust spectra, and the black curve is their sum. The top row represents the best fit obtained when the silicate and AC dust temperatures were unconstrained. The second and third row represent the best fits when the ratio of the AC-to-silicate dust temperatures were constrained to be 1.3 and 2.0, respectively. The bottom figure is the fit to the far-IR spectrum for pure AC dust. More details are in the text. {\bf Right column}: Map of \chisq\ as a function of the silicate and AC dust masses. The contour line represents the value of 1.5 times the minimum \chisq, and the X marks its minimum value. The corresponding dust temperatures and masses are given in Table~\ref{results1}.\label{specfits}} 
 \end{figure*}

 \section{The Dust  and Stellar Contributions to the galaxy's SED}
Figure~\ref{clashsed} presents the observed flux densities from the source, which are tabulated and referenced in Table~\ref{fluxes}. The spectral energy distribution comprises of three distinct contributions: thermal emission from dust at wavelengths above $\sim 50$~\mic;  an obscured stellar population from 0.4 to 2~\mic; and a population of unobscured young blue stars from $\sim 0.2$ to 0.4~\mic. The fits to the dust and stellar components of the SED are described in the following subsections.  
  
\subsection{The Dust Emission Component}
We assumed that the mid-IR to submm emission arises from astronomical silicate and amorphous carbon (AC) grains, with optical constants given by \cite{zubko04}. Model parameters were the silicate and AC dust masses, \msil\ and \mcrb, and their respective temperatures, \tsil\ and \tcrb. We varied dust masses from $1\times10^7$ to $1\times10^9$~\msun, and solved for the dust temperatures that gave the best fits to observed IR fluxes, ignoring upper limits.

The results of the fit are given in Figure~\ref{specfits}. The left column presents the fits to the SED, and the top left panel presents an unconstrained fit to the dust spectrum with silicate and AC dust masses and temperatures as free parameters. The mid-- to far--IR spectra are readily fit with 30~K silicate dust, however the steep $\lambda^{-2}$ falloff in the silicate emissivity failed to fit the millimeter fluxes. These were fit with a very cold, $\sim 7$~K, AC dust component. We consider this fit to be physically unrealistic, because of the large $\gtrsim 10^9$~\msun\ of AC dust needed to fit the spectrum, and the large disparity between the silicate and AC dust temperatures. Exposed to a radiation field similar to that in the diffuse ISM of the Galaxy, AC dust will attain a somewhat higher temperature than that of the silicates, with a ratio of \tcrb/\tsil $\approx 1.3$ \citep{zubko04}. We therefore calculated the best fits to the far--IR mm spectrum, constraining the AC-to-silicate dust temperature to be 1.3. To test the dependence of the derived dust masses and temperatures on this ratio, we also considered a dust model with a \tcrb/\tsil\ ratio of 2.0. The resulting fits are shown in the left column of the second and third rows of the figure. 
The left panel in the last row presents the fit to the spectrum with a single AC dust component. 

The right column in the figure shows the maps (figure) of \chisq\ as a function of dust masses (mass). When the temperatures of the different dust components are unconstrained (top row of the figure), the best fit is obtained with hot silicate and very cold  AC dust, radiating close to the lower limit of $\sim 6$~K imposed by the cosmic microwave background at that redshift. 
When the relative dust temperatures are constrained to more realistic values (second and third rows), the \chisq\ maps show that the AC dust mass is  well determined to be around $5.0\times 10^7$~\msun, whereas the silicate mass can vary substantially without significantly affecting the goodness of the fit to the data. 
This is a direct result of the temperature constraints and the differences between the spectral behavior of the emissivity of the two dust components. Because of its higher temperature, the AC dust provides a good fit to the SED at the shortest IR wavelengths, and dominates the emission at all other wavelengths. Consequently its mass is well defined. The emission from the silicate dust component is only a secondary contributor to the observed SED, and its temperature decreases with increasing AC dust temperatures, in order to fit the long wavelength emission. The resulting 30\% variation in silicate temperatures causes the mass of the silicate dust to vary by factors of 5. In contrast, there is very little variation in the AC dust temperature ($\lesssim 8$\%) resulting in less than a factor of 1.4 variation in their dust mass.   

Overall, the minimum value of \chisq\ for all cases is around 16 with 3 degrees of freedom, which is formally a poor fit to the data. This is largely due to the fact that the 1.1 and 1.3~mm data points are inconsistent with a declining spectrum, and that the GISMO~2~mm flux may contain the contribution of other galaxies from within the large beam. Eliminating the 1.1 and 1.3~mm data points will reduce the value of \chisq\ to about 1.8 for 1 degree of freedom, which is a marginally acceptable fit to the data.

Dust masses, temperatures, and luminosities are presented in Table~\ref{results1}. 
\subsection{The Stellar Emission Component}
To model the stellar emission component we adopted a star formation history (SFH) characterized by a delayed exponential function with a characteristic timescale $\tau$ of the form: $\psi(t)\propto ({t/\tau})\times \exp(-t/\tau)$ which, normalized to a SFR of $\psi_0$ at time $t_0$, can be written as:
\begin{equation}
\label{sfh}
\psi(t)=\psi_0\, \left({t\over t_0}\right)\times \exp[-(t-t_0)/\tau]
\end{equation}
where $\psi_0$ is the SFR at time $t_0=6$~Gyr, the age of \clash\ at $z=1$.

We used the population synthesis code PEGAS\'E \citep{fioc97} with a Kroupa IMF \citep{kroupa01} to calculate the intrinsic stellar spectrum, and adopted the Calzetti attenuation law \citep{calzetti00} to calculate the stellar radiation that escapes the galaxy, and the radiation that is absorbed by the dust. The current SFR was determined from the total bolometric luminosity, and was found to be equal to $\psi_0=56$\mul~\sfr. A good fit to the observed stellar component was  obtained for values of $\tau=6$~Gyr, and $A(V)=3.5$. The values of the intrinsic stellar luminosity, $L_{\star}(int)$, the escaping stellar luminosity, $L_{\star}(esc)$, and the stellar luminosity absorbed by the dust, $L_{\star}(abs)$, are listed in Table~\ref{galx}. The value of the absorbed stellar luminosity, $7.7\times 10^{11}$\mul~\lsun, is somewhat larger than the calculated IR luminosity, which is about $6.3\times 10^{11}$\mul~\lsun.
The fractional difference of 18\% is consistent with the missing luminosity from the mid--IR region, if \clash\ had an Arp220--like SED. The warm dust that emits  at mid--IR wavelengths makes a small contribution to the total IR luminosity, and a negligible one to the dust mass. The stellar emission  component is therefore consistent with the dust emission component. 

The rising stellar SED in the $\sim 0.2-0.4$~\mic\ region was modeled by a 100~Myr burst, with a SFR of 4~\sfr. The total burst luminosity and stellar mass are also listed in Table~\ref{galx}.

The specific SFR (sSFR), defined as $\psi_0/M_{\star}$, is equal to $4.3\times 10^{-10}$~yr$^{-1}$, or equivalently, the specific timescale for star formation, $\tau_{\star}$, defined as the inverse quantity,  is equal to 2.3~Gyr. The value of the sSFR is larger than that expected from a galaxy with a stellar mass of $\sim 1.2\times 10^{11}$\mul~\msun\ \citep{bauer05}. Alternatively, the value of $\tau_{\star}$ is much shorter than the age of the galaxy, suggesting that most stars formed in a recent episode of star formation.  

\section{The Equilibrium Dust Mass in Galaxies}

The mass of dust in galaxies is determined by the rates of dust injection into the ISM by AGB stars and core collapse supernovae (CCSN), and by their destruction rate by supernova remnants (SNRs)in the interstellar medium (ISM), and can be calculated by detailed evolutionary models, such as those presented by \cite{dwek98c} and \citep{dwek11a}.
If the SFR has been constant over the lifetime of the dust, $\tau_d(t)$, then the equilibrium dust mass dust mass is given by:
\begin{equation}
\label{mdust1}
M_d(t) = \sum_j\ R_j(t)\, Y_j\, \tau_d(t)
\end{equation}
where $R_j(t)$ is death rate of the dust producing sources $j\equiv$ \{AGB,CCSN\}, and $Y_j$ is their average dust yield. The dust lifetime is given by \citep{dwek80a}:
\begin{equation}
\label{taud}
\tau_d (t)= {\misme(t)\over \rsne(t)\, m_g}
\end{equation}
where \mism\ is the total mass of ISM gas, $m_g$ is the effective mass of gas that is completely cleared of dust by a single SNR, and \rsn\ is the rate of core collapse and Type~Ia SNe. 
The equilibrium dust mass can then be explicitly written as:
\begin{equation}
\label{mdust2}
M_d = \left[ \left({\rccsne\over \rsne}\right)\, Y_{\rm CCSN} + \left({\ragbe\over \rsne}\right)\, Y_{\rm AGB}\right]\, \left({\misme \over m_g}\right)
\end{equation}
The $R_j/$\rsn\ ratios depend only on the stellar IMF so that given a dust yield in the different stellar sources, the equilibrium dust mass primarily depends on the available ISM mass, and the mass of gas cleared of dust by each SNR.

\section{The Origin of the Dust in \clash}
The ratio of Type Ia to CCSN in the solar neighborhood is about 0.20 \citep{tammann94,cappellaro96} giving an \rccsn/\rsn\ ratio of 0.8. For a Kroupa IMF, the death rate of low mass carbon stars is 6 times higher than the total SN rate, giving \ragb/\rsn\ $\approx 6$. 

For simplicity we will assume that all the silicate dust is produced in CCSN with average yields of 0.1~\msun, based on observations of the Crab Nebula and Cas~A \citep{temim13,arendt14}. Carbon dust is made in the outflows from AGB stars in the 1.4 to 4.0~\msun\ mass range, with average yields of $(0.5-1)\times 10^{-2}$~\msun \citep{ferrarotti06,zhukovska08a,nanni13}.

The value of $m_g$ was recently calculated for a range of ISM densities, interstellar magnetic field intensities, and SN explosion energies, and found to be $(1-2)\times 10^3$~\msun, depending on these values \citep{slavin15}. 

So the equilibrium dust mass in galaxies is approximately given by:
\begin{equation}
\label{mdust3}
M_d^{eq} \approx 10^{-4}\, \misme
\end{equation}

If we use an ISM mass of $\sim 5\times 10^9$~\msun\ for the Milky Way, then the equilibrium dust mass is only $\sim 5\times 10^5$~\msun, considerably lower than the $\sim 3\times 10^7$~\msun\ inferred from dust models \citep[e.g.][]{zubko04}. This discrepancy reflects the fact we only considered stellar sources of interstellar dust, while detailed chemical evolution models suggest that  most of the Galactic dust must have been reconstituted in the dense ISM \citep{dwek80b,dwek98c,zhukovska07,calura10}. 

The ISM mass in \clash\ is unknown, but if we adopt the stellar mass of  $1.3\times 10^{11}$~\msun\ as an upper limit on the gas mass, then the upper limit on the equilibrium dust mass becomes $\sim 10^7$\mul~\msun, a factor of at least 5 times lower than that derived from fitting the far-IR emission, independent of the lensing magnification. 
The underlying reason for this discrepancy is that the dust production rates in stellar sources (CCSNe and AGB stars) are significantly lower than their destruction rates by SNR in the ISM.  The situation in \clash\ is therefore similar to that in the Milky Way, and the Magellanic Clouds and other external galaxies, in which the dust production rates are significantly lower than their destruction rates \citep{dwek98c,dwek07b,dwek11a,gall11c, valiante11, michalowski10b, slavin15, temim15, rowlands14, michalowski15}. A significantly lower destruction efficiency is required to balance the dust production and destruction rates in the ISM. Alternatively, grain growth by accretion onto the cores of surviving grains, as originally suggested for the Milky Way \citep{dwek80b,draine09}, may be an additional source of interstellar dust in \clash. 
 
\section{Summary}
\clash\ is a star forming galaxy at $z=1.0$ with a current SFR of 54\mul~\sfr, a total stellar mass of $1.3\times 10^{11}$\mul~\msun, a sSFR of $3\times 10^{-10}$~yr$^{-1}$, and a total dust mass of $\sim 5\times 10^7$\mul~\msun, where the mean value of $\mu$ is 2.7. The inferred dust mass  is  higher than the upper limit on the equilibrium dust mass from stellar sources.  Similarly to many other star forming galaxies, the dust mass in \clash\ cannot be accounted for by formation in CCSN and AGB stars, requiring the need to grow most of the dust in the dense phases of the ISM.

\acknowledgements
This work was supported by NASA's 12-ADP12-0145 and 13-ADAP13-0094 research grants, and supported through NSF ATI grant 1106284. IRAM is supported by INSU/CNRS (France), MPG (Germany) and IGN (Spain). ST acknowledges support from the ERC Consolidator Grant funding scheme (project ConTExt, grant number. 648179). This work utilizes gravitational lensing models produced by PIs Brada$\breve{c}$, Ebeling, Merten \& Zitrin, Sharon, and Williams funded as part of the HST Frontier Fields program conducted by STScI. STScI is operated by the Association of Universities for Research in Astronomy, Inc. under NASA contract NAS 5-26555. The lens models were obtained from the Mikulski Archive for Space Telescopes (MAST), https://archive.stsci.edu/prepds/frontier/lensmodels/.  

\begin{deluxetable}{lll}
\tablewidth{0pt}
\tablecaption{Observed fluxes from \clash.}
\tablehead{
\colhead{Wavelength (\mic)} &
\colhead{Flux (mJy)} &
\colhead{Reference} 
}
\startdata
     0.237  &   0.00091 $\pm$ 0.00014  & 0 \\
     0.271  &   0.00217   $\pm$  0.00015  & 0 \\
     0.336  &   0.00184   $\pm$  0.00012  & 0 \\
     0.392  &   0.00185   $\pm$  0.00007  & 0 \\
     0.436  &   0.00224   $\pm$  0.00008  & 0 \\
     0.442  &   0.00188   $\pm$  0.00001  & 1 \\
     0.478  &   0.00225   $\pm$  0.00007  & 0 \\
     0.540  &   0.00297   $\pm$  0.00001  & 1 \\
     0.541  &   0.00275   $\pm$  0.00005  & 0 \\
     0.596  &   0.00364   $\pm$  0.00005  & 0 \\
     0.632  &   0.00434  $\pm$  0.00008  & 0 \\
     0.647  &   0.00487   $\pm$  0.00002  & 1 \\
     0.776  &   0.00909   $\pm$  0.00010  & 0 \\
     0.810  &   0.01000   $\pm$  0.00004  & 0 \\
     0.895  &   0.01490   $\pm$  0.00013  & 0 \\
     0.905  &   0.01562   $\pm$  0.00004  & 1 \\
     1.055  &   0.01810   $\pm$  0.00007  & 0 \\
     1.153  &   0.02110   $\pm$  0.00006  & 0 \\
     1.249  &   0.02390   $\pm$  0.00008  & 0 \\
     1.392  &   0.02770   $\pm$  0.00007  & 0 \\
     1.537  &   0.03100   $\pm$  0.00007  & 0 \\
     2.150  &   0.06199   $\pm$  0.00032  & 1 \\
     3.553  &   0.08929   $\pm$  0.02232  & 2 \\
     4.449  &   0.07210   $\pm$  0.01803  & 2 \\
   100    & 7.7 $\pm$1.0  & 2  \\
   160    & 18.4 $\pm$ 1.9  & 2  \\
   250    & 23.7 $\pm$ 3.7  & 2  \\
   350    & 14.2 $\pm$ 3.0  & 2  \\
   500    & 6.1  $\pm$ 3.9  & 2  \\
   102.259  &   9.04997   $\pm$  2.26249  & 2 \\
   163.842  &  19.22050   $\pm$  4.80513  & 2 \\
   252.780  &  22.01670   $\pm$  5.50417  & 2 \\
   356.183  &  17.35570   $\pm$  4.33893  & 2 \\
  1100.00   &   0.62     $\pm$ 0.17  & 7 \\
   1293.880  &   1.232   $\pm$  0.255  & 5 \\ 
  2000.000  &   0.40000   $\pm$  0.09800  & 6 
\enddata
\label{fluxes}
\tablenotetext{1}{References: 
(0) HST: CLASH HST Catalog = source 2882 \citep{postman12};
(1) Subaru: CLASH Subaru Catalog = source 61021;
(2) Rawle et al. (2015, in preparation)
(3) WISE: AllWISE Source Catalog = source J114934.41+222442.5 (confused!);
(4) SHARC2: Dwek et al. (2014);
(5) NOEMA: (this work);
(6) GISMO: Dwek et al. (2014);
(7) AzTEC: Zavala et al. (2015)
}
\end{deluxetable}

\begin{deluxetable}{lcccc}
\tablewidth{0pt}
\tablewidth{0pt}
\tablecaption{Best fit dust parameters \tablenotemark{1}}
\tablehead{
\colhead{$T_{crb}/T_{sil}$ constraint} &
\colhead{none  } &
\colhead{1.3} &
\colhead{2.0} &
\colhead{AC dust} 
}
\startdata
$M_{sil}$ ($10^8$~\msun)     & 2.1           & $<0.1$         & 0.26           & \nodata  \\
$T_{sil}$     (K)            & 30.4$\pm$0.24 & 29.3$\pm$0.3   & 19.2$\pm$0.2   & \nodata  \\
$L_{sil}$ ($10^{11}$~\lsun)  & 6.0           & 0.23           & 0.045          & \nodata \\
\hline
$M_{crb}$ ($10^8$~\msun)     & $>10$         & 0.50           & 0.50          & 0.52  \\
$T_{crb}$ (K)                & 7.3$\pm$0.5  & 38.1            & 38.3          & 38.0  \\
$L_{crb}$ ($10^{11}$~\lsun)  & 0.03           & 6.0           & 6.2           & 6.3  \\
\hline
$M_{dust}$ ($10^8$~\msun)    & 2.1           & 0.50           & 0.76           & 0.52  \\
$L_{dust}$ ($10^{11}$~\lsun) & 6.0           & 6.2            & 6.3           & 6.3   
\enddata
\label{results1}
\tablenotetext{1}{Dust masses and luminosities were not corrected for lensing amplification.}
\end{deluxetable}

\begin{deluxetable}{lll}
\tablewidth{0pt}
\tablecaption{Best fit galactic parameters\tablenotemark{1},\tablenotemark{2}}
\tablehead{
\colhead{Quantity} &
\colhead{ } &
\colhead{Value} 
}
\startdata
Continuous star formation & & \\
\hline
$\psi_0$~(\sfr)  & &54 \\
$\tau$ (Gyr) & &6\\
$A(V)$ & & 3.5 \\
$\tau(V)$ & & 1.44 \\
$M_{\star}$ (\msun) & &$1.3\times 10^{11}$ \\
sSFR $yr^{-1}$ & & $4.3\times 10^{-10}$ \\
$\tau_{\star}$ (Gyr) & &2.3 \\
$L_{\star}(int)$ (\lsun) & &$9.3\times 10^{11}$ \\
$L_{\star}$(esc) (\lsun) & &$1.6\times 10^{11}$ \\
$L_{\star}$(abs) (\lsun) & &$7.7\times 10^{11}$ \\
$L_{dust}$ (\lsun) & &$6.3\times 10^{11}$ \\
\hline 
Burst of star formation & & \\
\hline
$\psi_0$~(\sfr)  & &4 \\
$L_{\star}$ (\lsun) & &$1.1\times 10^{10}$ \\
$M_{\star}$ (\msun) & &$3.2\times 10^8$ 
\enddata
\label{galx}
\tablenotetext{1}{calculations were done for a Kroupa IMF}
\tablenotetext{2}{Star formation rates, luminosities, and stellar masses were not corrected for lensing amplification.}
\end{deluxetable}


\clearpage

 \end{document}